\documentclass[aps,prb,twocolumn,groupedaddress,showpacs,floatfix]{revtex4}

\usepackage{graphicx}

\newcommand{\re}{{\bf r}  }

\begin{document}

\title{Photoelectron spectra of anionic sodium clusters from time-dependent
  density-functional theory in real-time}

\author{Michael Mundt and Stephan K\"ummel}
\affiliation{Institut f\"ur Theoretische Physik, Universit\"at Bayreuth,
D-95440 Bayreuth, Germany}

\date{\today}

\begin{abstract}
We calculate the excitation energies of small neutral sodium clusters in the
framework of time-dependent density-functional theory. In the presented
calculations, we extract these energies from the power spectra of the dipole
and quadrupole signals that result from a real-time and real-space
propagation. For comparison with measured photoelectron spectra, we use the
ionic configurations of the corresponding single-charged anions. Our
calculations clearly improve upon earlier results for photoelectron spectra
obtained from static Kohn-Sham eigenvalues.
\end{abstract}


\maketitle

\section{Introduction}

Since more than one hundred years, photoelectron spectroscopy plays an
important role in physics. Einstein's explanation of the photo-effect,
probably the most well-known experiment in the field, was a crucial step in
the development of quantum mechanics. Whereas the photo-effect revolutionized
the understanding of light, the main aim of modern photoelectron spectroscopy
is to understand electronic and ionic structures from solids down to single
atoms. Especially in the context of nanoscale materials photoelectron
spectroscopy is one of the most important experimental tools since it is
almost the only method that provides access to the electronic and ionic
structures of these materials. The direct observation of the electronic shell
structure in sodium clusters \cite{wrigge} is just one example for the power
of the method. Another application is the determination of the ionic structure
of, e.g., clusters. Since the electronic structure, and thus the photoelectron
spectrum (PES), depends on the ionic configuration, comparing the measured PES
with the results from first-principle calculations allows the identification
of the ionic structure. This interplay between theory and experiment has
already been used successfully in many cases
\cite{gantefoer1,akolaal,jenadetach,kronik,mosnegna,gantefoer2,jyvasna59,
agaucupes,Kost07}.

Clearly, the just mentioned method can only work if reliable calculations for
the system of interest can be performed. Since most of the measured systems
consist of many electrons, density-functional theory (DFT) is an especially
well-suited tool due to its low numerical costs. Unfortunately, evaluating the
PES from a Kohn-Sham (KS) DFT calculation is not an easy task since only the
highest occupied KS eigenvalue has a rigorous connection to the PES: it is
equal to the ionization potential \cite{ionizationupot, almbladh,
PerdLevComm}. Thus, it yields the position of the first peak in the PES.

The most common approach to obtain the other peaks in the PES from a DFT
calculation is based on the density of states of the occupied KS orbitals. In
this approach the KS eigenvalue spectrum of the `mother' system, i.e., the
system which still contains the photoelectron, is directly compared to the
experimental PES. In many situations the resulting spectra reproduce the
experimental ones quite well \cite{gantefoer1,akolaal,jenadetach,kronik,
mosnegna,gantefoer2,jyvasna59,agaucupes,chong,MWHH1,Kost07}.

Another way to extract the information related to the PES from a DFT
calculation is via the excitation energies of the `daughter' system, i.e., the
system with one electron less. Since time-dependent DFT (TDDFT) \cite{RG,
TDDFTRev}, in principle, allows to calculate the excitation energies of a
system exactly, TDDFT can be used to calculate the positions of the PES peaks
accurately. This approach is followed in Ref.\ \onlinecite{C84KA} and Ref.\
\onlinecite{MWHH2}. The basic idea of this approach is also used in Ref.\
\onlinecite{BonKou}, but in combination with configuration-interaction and not
TDDFT calculations.

We finally want to mention a third method how the PES can be obtained from a
DFT calculation. In this approach, the time-dependent ionization process is
simulated in real-time and the kinetic energy spectrum of the outgoing
components of the KS Slater determinant are connected to the PES \cite{APPG2}.

In Ref.\ \onlinecite{mosnegna}, the PES resulting from Na$_5^-$, Na$_7^-$, and
Na$_9^-$ (among others) irradiated by an XeCl excimer laser ($\hbar \omega =
4.02$ eV) were measured and compared with the corresponding KS eigenvalue
spectra. Although the agreement between the theoretical results and the
measured PES was generally reasonable, a systematic discrepancy was
found. Namely, the width of the theoretical spectrum, i.e., the difference
between the energy of the energetically highest and lowest occupied KS
eigenvalue, was too large by about 0.2-0.4 eV. In Ref.\ \onlinecite{mmsk} the
reason for this discrepancy was examined. It was shown that technical aspects,
e.g., the treatment of the pseudopotential, could not explain the
differences. Furthermore, it was demonstrated that using the exchange-only
optimized-effective potential (OEP) \cite{OEP1} reduced the width of the KS
spectrum but not to an extent that would bring the spectrum in agreement with
experiment.  In addition, it was also shown that using Slater's transition
state concept could also not improve the theoretical results.  Thus, the
question arises whether a different method to extract the PES from a DFT
calculation leads to a better agreement with the experiment. It is the aim of
the present manuscript to answer this question by extracting the PES from the
excitation energies of the corresponding `daughter' systems.

\section{Theoretical background}

Before discussing the results let us sketch the theoretical background of the
presented calculations in more detail.
\begin{figure}[t]
\includegraphics[clip,width=8.1cm]{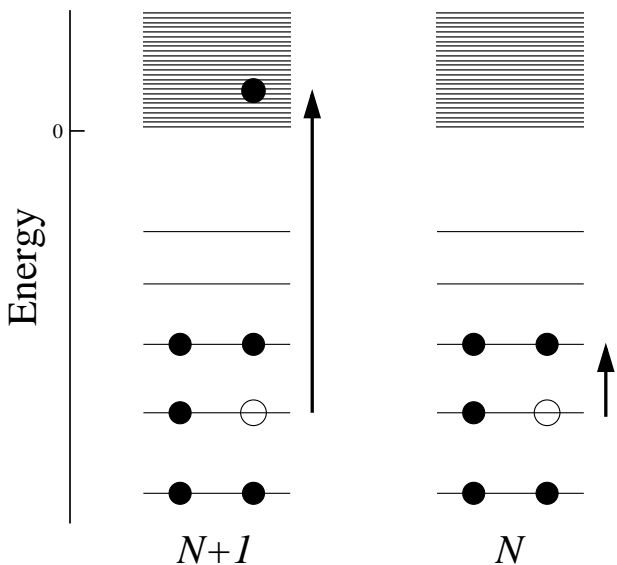}
\caption{Schematic view of two different approaches to calculate the
  PES. Left: The process is described as a strong excitation of the
  $(N+1)$-electron system. Right: The photoelectron has already been detected
  and the remaining $N$-electron system is left in an excited state. The link
  between the kinetic energy of the photoelectron and the energy of the
  excited state of the $N$-electron system is provided by energy conservation.
\label{Fig:PESschematic}}
\end{figure}
Fig.\ \ref{Fig:PESschematic} schematically shows two approaches how the peak
positions in the PES can be calculated. On the left hand side, the process is
described as an excitation process from the ground state to an energetically
high-lying state with continuum contributions. Since KS eigenvalue differences
are zeroth-order approximations to excitation energies \cite{AGExcitKS, Gonze},
the KS DOS of the $(N+1)$-electron system can be used to obtain an approximate
PES. In addition to this argument, Chong {\it et al.} \cite{chong} have given
well founded arguments that KS eigenvalues can be interpreted as
approximations to relaxed vertical ionization potentials.

On the right hand side of Fig.\ \ref{Fig:PESschematic}, the situation after
the photoelectron has been detected is considered. In this case, the remaining
system is left in an energetically low-lying excited state of the $N$-electron
system. To connect the excitation energies of this system to the PES, energy
conservation is used. Before the photon is absorbed the total energy is given
by $E_0^{(N+1)} \, + \, \hbar \omega$, where $E_0^{(N+1)}$ is the ground-state
energy of the `mother' system containing $N+1$ electrons and $\hbar \omega$ is
the photon energy. After the detection of the photoelectron the total energy
is given by the kinetic energy of the photoelectron $E_{\mathrm{kin}}$ and the
energy of the remaining `daughter' system. Since the total energy is conserved,
it follows that
\begin{eqnarray}
E_{\mathrm{bind},j} = E_{\mathrm{kin}} - \hbar \omega = E_0^{(N+1)} -
E_0^{(N)} - \Delta E_j^{(N)}
\label{Eq:Econst}
\end{eqnarray}
must hold. Here, $E_0^{(N)}$ is the ground-state energy of the `daughter'
system and $\Delta E_j^{(N)}$ are its excitation energies \cite{Remark1}. For
the first peak in the PES the kinetic energy of the photoelectron is
maximal. In this case, the `daughter' system is in its ground-state, i.e.,
$\Delta E_j^{(N)}$ is zero and the peak position is at $E_0^{(N+1)} -
E_0^{(N)}$.

To obtain the excitation energies from time-dependent DFT the full linear
density-response function of the interacting system can be used. This function
provides access to the excitation energies of the system since it has poles at
these energies. The crucial observation now is that the interacting linear
density-response function can be expressed in terms of the KS response
function and the exchange-correlation (xc) kernel \cite{TDDFTExcit, Casida,
TDDFTRev}. Nowadays, most applications use the matrix equation of Casida
\cite{Casida} to obtain these excitation energies.

Alternatively, the excitation energies can be extracted from a spectral
analysis of the time-dependent density coming from a real-time propagation
\cite{Bertsch,PG,Octop}. In this approach, the xc kernel is not needed, but
instead, the time-dependent KS equations are solved without explicit
linearization.  To illustrate this approach, imagine we have created a
time-dependent density $n(\re ,t)$ of an interacting system by, e.g., a laser
excitation. Assuming that the system is confined by the same time-independent
potential before and after the laser pulse we can write the excited density in
terms of the eigenstates $| \psi_j \rangle$ of the interacting system in the
time-independent potential. It reads
\begin{eqnarray}
n(\re ,t) &=& \langle \psi (t) | \hat{n} | \psi(t) \rangle \nonumber \\ &=&
          \sum_{j,k} c_j^* c_k \langle \psi_j | \hat{n} | \psi_k \rangle
          \exp(-i(E_k-E_j)t/\hbar)
\label{Eq:DensExp}
\end{eqnarray}
where $E_j$ is the eigenvalue corresponding to $| \psi_j \rangle$ and
$\hat{n}$ is the density operator. Assuming that the time-dependent state
$|\psi(t)\rangle$ is dominated by the ground state, i.e., $c_0 \gg c_j$ we can
write
\begin{eqnarray}
n(\re ,t) &\approx& |c_0|^2 n_0(\re) \ + \nonumber \\ &\sum_{j}& c_0^* c_j
 \langle \psi_0 | \hat{n} | \psi_j \rangle \exp(-i(E_j-E_0)t/\hbar) + c.c. \
 ~~~
\label{Eq:Density}
\end{eqnarray}
Here, $n_0(\re)$ is the ground-state density of the system. If we now
calculate the Fourier transform of $n(\re ,t)$ we will get peaks at the exact
excitation energies of the system. Since time-dependent DFT in principle
provides us with the exact time-dependent density, this is an easy method to
obtain the excitation energies of the interacting system from a time-dependent
DFT calculation.

In a practical calculation, two problems must be solved to get the excitation
energies from this scheme. First, one has to create a time-dependent density
which is dominated by the ground-state density and, in addition, contains the
excited states of interest.  The second problem is how to extract the
excitation energies from the time-dependent density in practice.  Since the
density in every space point at all times cannot be stored, a full Fourier
transform of Eq.\ (\ref{Eq:Density}) giving $n(\re ,\omega)$ is not
possible. To overcome this problem several possibilities exist. One is to
evaluate $n(\re ,\omega)$ only for some points in space \cite{SKPGDensPlas},
e.g., in the center of the cluster.  A different method is to Fourier
transform certain moments of the density distribution. Typically, the dipole
moment is used for this purpose \cite{Bertsch,PG}. Obviously, some excitations
are filtered out by this procedure because the Fourier spectrum of the dipole
moment only shows excitation energies of states which can be coupled to the
ground state via the dipole operator. Non-dipole active excitations can be
taken into account by recording higher moments. In the following we will use
the time-dependent dipole and quadrupole moments obtained from a real-time
propagation to obtain excitation energies for the systems of interest.

\section{Technical aspects}

For the ionic ground-state configurations of the `mother' systems we used
optimized structures obtained with the PARSEC \cite{Parsec} program
package. The generalized-gradient approximation of Perdew {\it et al.} (PBE)
\cite{PBE} was employed for the geometry optimization. The ionic cores were
treated consistently with norm conserving non-local pseudopotentials
\cite{TroullierM}.

The time-dependent KS equations were solved on a real-space grid in real
time. For this we implemented the necessary algorithms into a modified version
of the PARSEC code. In detail, we implemented a fourth-order Taylor
approximation to the propagator in combination with a higher-order
finite-difference formula for the kinetic part of the KS Hamiltonian. For the
calculations, we used a time step of 0.003 fs and the total propagation time
was 75 fs. The ionic cores were again described by norm conserving non-local
pseudopotentials. Furthermore, the ionic structures were fixed during the
propagation. The grid spacing was 0.7 $a_0$ and the grid radius varied between
20 and 23 $a_0$ depending on the system. The time-dependent density was
created by applying a boost $\exp (i \, \mathbf{r} \cdot
\mathbf{p}_{\mathrm{boost}} / \hbar)$ to the ground-state KS orbitals.  The
total excitation energy of the system was $E_{\mathrm{excit}} = 1.0 \times
10^{-5}$ eV, i.e., a boost strength $|p_{\mathrm{boost}}| = \sqrt{2 m_e
E_{\mathrm{excit}} / N}$ was applied to each KS orbital (with $m_e$ being the
electron mass and $N$ the number of electrons). In addition, the calculations
were repeated with a boost strength reduced by a factor of $1.0 \times
10^{-2}$.  Using these two small boost strengths allowed us to check whether
the created time-dependent density was dominated by the ground-state density
(see below).

Instead of applying the same boost vector $\mathbf{p}_{\mathrm{boost}}$ to all
KS orbitals, and thus creating a coherent velocity field, we varied the boost
direction for different KS orbitals. This is necessary since applying the same
boost direction to all KS orbitals corresponds to first order in
$\mathbf{p}_{\mathrm{boost}}$ to a dipole excitation of the system, i.e., from
the resulting time-dependent density it is only possible to retrieve the
excitation energies of `dipole-active' states.  By applying different boost
directions to different KS orbitals we modeled a general excitation mechanism
creating a time-dependent density containing excited states with different
symmetry properties.  In detail, we randomly chose a boost direction (no
symmetry axis of the considered cluster) for the first orbital and then chose
our coordinate system such that this direction was the first diagonal (for the
remaining rotational degree of freedom a random angle was chosen). After this
we boosted the second orbital in the opposite direction of the first
boost. The third orbital was then boosted in the direction of the second
diagonal of the chosen coordinate system, the forth again in the opposite
direction and so on. For Na$_9$, the ninth orbital was boosted again in the
same direction as the first orbital. Since the only purpose of this procedure
was to create a time-dependent density without any particular symmetries, we
do not consider the relative orientation of the cluster with respect to the
boost directions to be of special importance.

Finally, we used the time-dependent local-density approximation (TDLDA)
\cite{ALDA} for the xc potential for the propagation. Since the linear
response of the homogeneous electron gas is the same in this approximation and
in the PBE functional, the differences in the resulting excitation energies
can be expected to be small in the low-energy regime. Unfortunately, the
exact-exchange orbital functional cannot be used for comparative calculations
since it requires a solution of the time-dependent optimized-effective
potential equation \cite{TDOEP} and no method for this exists in real-time at
the moment \cite{MMSKTDOEP}.

\section{Results and discussion}

\subsection{Results for Na$_3^-$}

\begin{figure}[t]
\includegraphics[width=8.1cm]{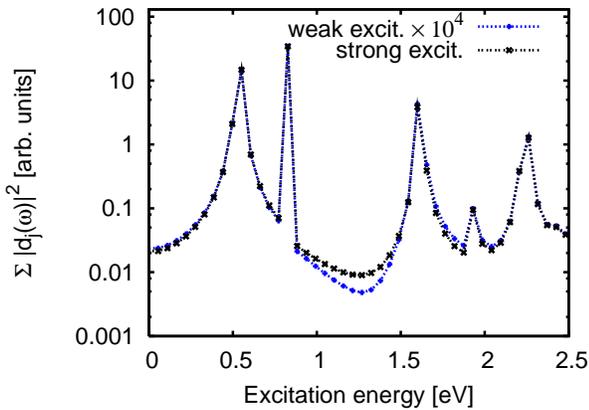}
\caption{(Color online) Dipole power spectrum of Na$_3$ resulting from an
  incoherent boost excitation. The result obtained from a total excitation of
  $1 \times 10^{-5}$ eV is labeled `strong excit.' whereas the label `weak
  excit.' corresponds to a boost reduced by a factor of $1 \times
  10^{-2}$. Clearly, the dipole power spectrum scales quadratically with the
  boost strength indicating that the peak positions correspond to excitation
  energies between the ground state and excited states.
\label{Fig:Na3DipSpec}}
\end{figure}

Fig.\ \ref{Fig:Na3DipSpec} shows the dipole power spectra of Na$_3$ resulting
from two boost strengths differing by a factor of $10^2$.  The dipole power
spectrum is given by
\begin{eqnarray}
D(\omega) \ := \ \sum_{j=1}^{3} |d_j(\omega)|^2
\end{eqnarray}
with $d_j(\omega)$ being the Fourier transform of the dipole moment
\begin{eqnarray}
d_j(t) \ = \ \int x_j \, n(\re,t) \ d^3r
\end{eqnarray}
where $x_1$ corresponds to the Cartesian coordinate $x$, $x_2$ to $y$, and
$x_3$ to $z$.
For small momentum boosts first-order perturbation theory predicts a linear
dependence of the expansion coefficients $c_j$ in Eq.\ (\ref{Eq:DensExp}) on
the boost strength. As a consequence, reducing the boost strength by a factor
of $c$ suppresses peaks corresponding to energy differences between two
excited eigenstates by a factor of $c^4$ in the power spectrum. Since peaks
corresponding to transitions between the ground state and an excited
eigenstate are only suppressed by a factor of $c^2$, changing the boost
strength allows one to distinguish between these two kinds of excitations. As
one can see in Fig.\ \ref{Fig:Na3DipSpec}, the results for the two boost
strengths are almost identical except for the predicted factor of
$10^4$. Thus, we conclude that all the peak positions in the dipole power
spectrum of Fig.\ \ref{Fig:Na3DipSpec} correspond to energy differences
between the ground state energy and the energy eigenvalues of the excited
eigenstates.

\begin{figure}[t]
\includegraphics[width=8.1cm]{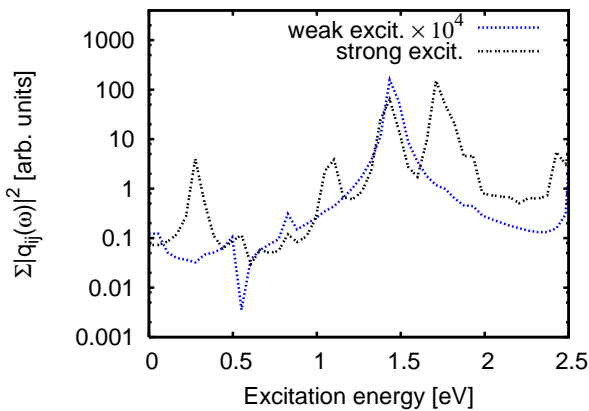}
\caption{(Color online) Sum of the absolute square of the Fourier-transformed
  components of the quadrupole tensor resulting from the same excitations as
  in Fig.\ \ref{Fig:Na3DipSpec}. In contrast to the dipole power spectrum some
  peaks vanish with reduced boost strength indicating that they correspond to
  energy differences between excited states.
\label{Fig:Na3QuadSpec}}
\end{figure}

The situation is different for the power spectrum resulting from the
quadrupole moments. In Fig.\ \ref{Fig:Na3QuadSpec} we plot
\begin{eqnarray}
Q(\omega) := \sum_{\stackrel{i=1}{j \ge i}}^3 |q_{ij}(\omega)|^2
\label{Eq:QuadSpecSum}
\end{eqnarray}
for the same two excitation boosts. In this equation, $q_{ij}(\omega)$ is the
Fourier transform of the quadrupole moment
\begin{eqnarray}
q_{ij}(t) &=& \int n(\re,t) \, (3 x_i x_j \, - \, r^2 \delta_{ij})\ d^3r \ ,
\\ r^2 &=& \sum_{i=1}^{3} x_i^2 \ , \nonumber
\label{Eq:QuadtenComp}
\end{eqnarray}
and the sum only runs over the independent components of the quadrupole
tensor.  Clearly, the quadrupole spectra for the different excitation
strengths differ considerably. For instance, the three large peaks at around
$0.3$, $1.1$, and $1.7$ eV vanish almost completely. Thus, we conclude that
they belong to transition energies between different excited states. Indeed,
one can see that these energies are exactly equal to the energy differences
between the first excited state and the other excited states from the dipole
spectrum.

The reason why there are no peaks at these energies in the dipole spectrum can
easily be understood if one takes the geometry of Na$_3$ into account. Since
Na$_3$ has a linear ionic configuration, the ground state has even
parity. Thus, the dipole spectrum only shows excited states with odd
parity. Since two states with odd parity cannot be coupled by the dipole
operator, transitions between these states do not show up in the dipole
spectrum.

\begin{figure}[t]
\includegraphics[width=8.1cm]{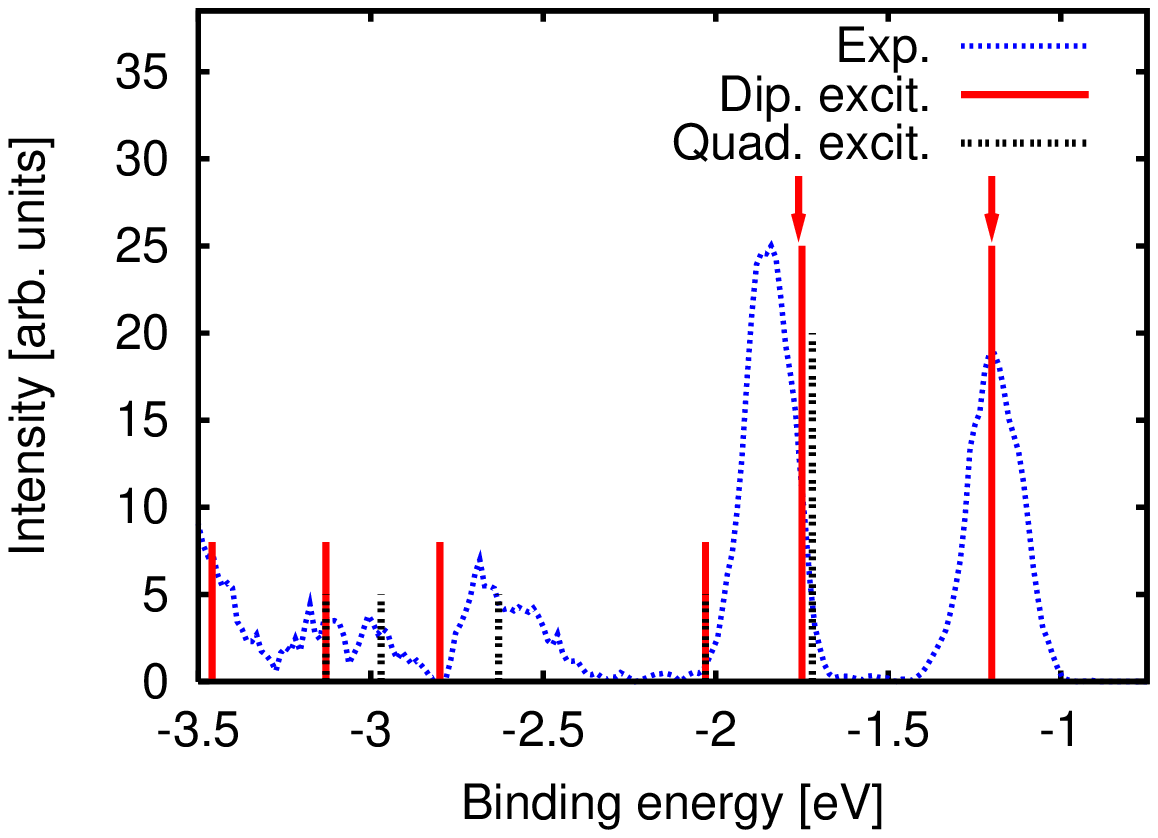}
\includegraphics[width=8.1cm]{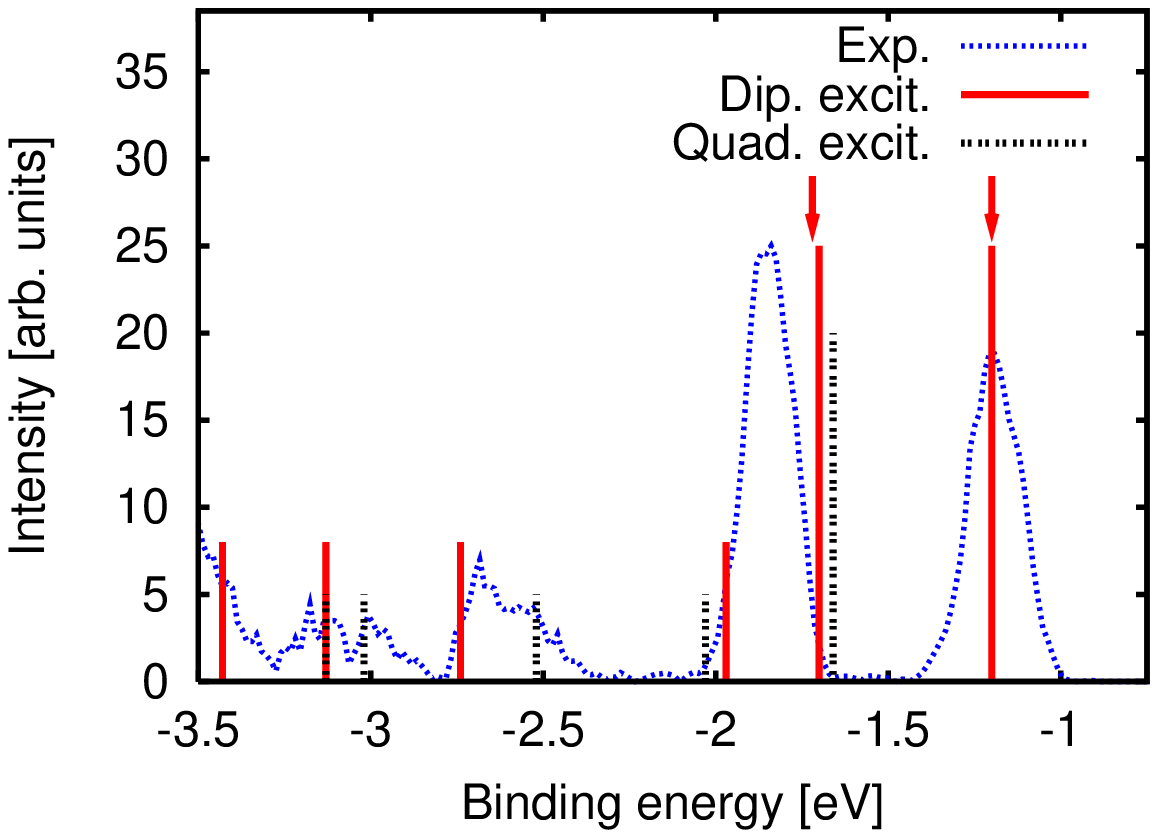}
\caption{(Color online) Measured PES of Na$_3^-$ (`Exp.') and theoretical PES
  obtained from the excitation energies of Na$_3$. Excitation energies from
  the dipole spectrum are labeled `Dip. excit.' whereas `Quad. excit.' labels
  excitation energies deduced from the quadrupole moments. Arrows indicate the
  result obtained from the KS DOS. Upper part: results obtained from the ionic
  ground-state configuration at zero temperature. Lower part: results obtained
  from an ionic configuration with a larger bond length to simulate a higher
  temperature. For most peaks the agreement with the experimental PES is
  clearly improved.
\label{Fig:1Na3Exp}}
\end{figure}

After the identification of the true excitation energies we can now compare
the results with the measured PES. In Fig.\ \ref{Fig:1Na3Exp} the excitation
energies of Na$_3$, the KS DOS of Na$_3^-$ and the measured PES (of Na$_3^-$)
are plotted. The positions of the occupied KS eigenvalues are indicated by
arrows, long bars indicate excitation energies from the dipole spectrum and
shorter bars excitation energies from the quadrupole moments.  In addition,
excitation energies leading to peaks below the strongest bound experimental
peak are reduced in their overall height. For better comparison, the KS DOS
and the excitation spectrum are both rigidly shifted in such a way that the
most weakly bound peak coincides with the experimental one.

As the upper part of Fig.\ \ref{Fig:1Na3Exp} shows the peak positions that one
obtains from the KS DOS are close to the experimental peak positions.  Unlike
in the case of larger Na clusters, the width is slightly smaller than the
energy difference between the two large experimental peaks but it is still
reasonable. However, since there are only two occupied KS orbitals in Na$_3^-$
the KS DOS picture fails completely to describe the higher lying peaks in the
measured spectrum.

As one expects from Eq.\ (\ref{Eq:Econst}) the PES obtained from the
excitation energies shows a much richer structure than the KS DOS. One
striking feature for instance is the second excitation around $-2.0$ eV. It
seems that the energy difference between this peak and the one at $-1.7$ eV is
too large in the calculation and that they are merged to one peak in the
experimental PES.  However, in general, the dynamically calculated excitation
energies and the energies obtained from the experimental PES are close to each
other even for the stronger bound peaks.  To see if the remaining discrepancy
can be further reduced by taking temperature effects into account we have
repeated our calculations with a larger bond length. Due to the net negative
charge of the cluster one can expect that other geometry changes, e.g.,
bending, only play a minor role in the case of Na$_3^-$. We have used a new
bond length of approx. $6.8$ instead of $6.5$ $a_0$. This new value for the
bond length $l$ of the cluster has been obtained from an estimate for the
thermal expansion at $T = 300$ K. It is based on the formula $\beta =
\frac{1}{l} \frac{\partial l}{\partial T}$ for the linear thermal expansion
coefficient $\beta$ which we have roughly estimated by $\beta \approx 2 \,
\beta_{\mathrm{bulk}}$ (see Ref.\ \onlinecite{SKTherm}), where
$\beta_{\mathrm{bulk}}$ is the bulk value for crystalline sodium at room
temperature.

The result can be seen in the lower part of Fig.\ \ref{Fig:1Na3Exp}.  For most
peaks one can observe a small shift towards lower absolute binding
energies. Except for the peak at $-1.7$ eV the agreement between the
experimental and theoretical spectrum is slightly improved by the increased
bond length. Especially the broader peak at around $-2.6$ eV is nicely
reproduced in this case. All in all, both calculations show that for
Na$_3^{-}$ the main advantage of the `excitation picture' is the reproduction
of the deeper bound structures in the PES.

\subsection{Results for Na$_5^-$}
\label{SubsecNa5}

\begin{figure}[t]
\includegraphics[width=8.1cm]{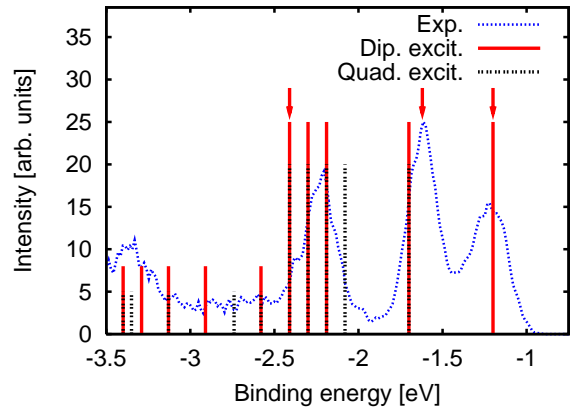}
\caption{(Color online) Same as in Fig.\ \ref{Fig:1Na3Exp} but for
  Na$_5^-$. Although both, the KS DOS and the PES from the excitation
  energies, describe the measured PES acceptable the large peak at $-2.2$ eV
  is much better reproduced by the excitation energies from the `daughter'
  system. The experimental data is taken from Ref.\ \onlinecite{mosnegna}.
\label{Fig:2Na5Exp}} 
\end{figure}

Fig.\ \ref{Fig:2Na5Exp} shows the experimental PES of Na$_5^-$, the KS DOS,
and the PES obtained from the excitation energies of Na$_5$. The labeling is
the same as in the corresponding previous figures. As for Na$_3^-$, the KS DOS
is in acceptable agreement with the first large peaks, although the strongest
bound large peak has a too negative binding energy in the KS DOS. As one can
see these peaks are also well described by the excitation energies of the
`daughter' system with the additional advantage that the last peak at $-2.2$
eV is better reproduced. In this approach it consists of four close-lying
excitations.

Beyond the peak at $-2.2$ eV the comparison with the experimental measurement
is difficult since no clear peak structures can be observed. Perhaps the
accumulation of excited states around $-3.3$ eV can be associated with the
measured peak in this region, but for the reasons given below, one has to be
very cautious in making comparisons in this part of the spectrum.

As one can see from the results for Na$_7^-$ and Na$_9^-$ discussed below the
problem of comparing the deeper lying part of the measured PES with calculated
excitation energies is not specific to Na$_5^-$. In general, the density of
excited states grows with the excitation energy, i.e., more and more states
appear in the theoretical calculation. On the other hand, from the point of
view of first-order perturbation theory the PES depends not only on the
positions of the excited states but also on the matrix element of the
perturbing operator $\hat{D}$ between the initial and the final state. Taking
the ground state of the `mother' system for the initial state and a product
state consisting of one photoelectron with momentum $\mathbf{k}$ for the final
state one obtains matrix elements of the form $\langle \mathbf{k}, \,
\psi_{j}^{(N)}| \ \hat{D} \ | \psi_{0}^{(N+1)} \rangle$. It is intuitively
clear that these matrix elements are much larger for low-lying states than for
energetically high-lying ones which in an independent-particle picture would
correspond to removing one particle and exciting a second one above the Fermi
level.  Especially, in the case of truly independent particles this process
cannot happen if the perturbing operator is a one-particle operator like the
dipole operator. Thus, many energetically high-lying eigenstates of the
`daughter' system are hardly or even not at all excited in the
experiment. Since the mentioned matrix elements depend on the interacting
many-particle wavefunctions, calculating these exactly is close to being
impossible. Especially, retrieving these matrix elements from a TDDFT
propagation of the `daughter' system is not trivial because the propagation
only provides information about matrix elements between excited states and the
$N$-particle ground state and not the $(N+1)$-particle ground state.

However, as the presented calculations show, the matrix elements do not play a
very important role in the part of the spectrum that we are mainly interested
in. Nevertheless, the calculations also clearly indicate that one has to
consider them if the deeper lying parts of the spectrum are of interest. A
possible method how this can be done in a TDDFT calculation can be found at
the end of Sec.\ \ref{Sec:SumCon}.

\subsection{Results for Na$_7^-$}

\begin{figure}[t]
\includegraphics[width=8.1cm]{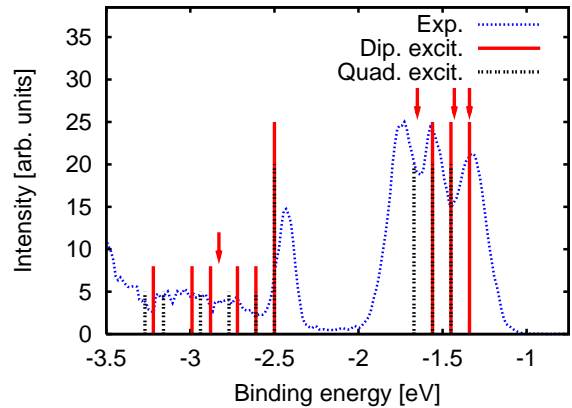}
\caption{(Color online) Same as in Fig.\ \ref{Fig:1Na3Exp} but for
  Na$_7^-$. Especially, the peak at $-2.4$ eV is much more accurately
  described by the excitation energies than by the KS DOS. In the weakly bound
  region thermal effects play an significant role in the case of
  Na$_7^-$. This explains the rather poor agreement between the theoretical
  values calculated at zero temperature and the measured curve between $-1.3$
  and $-1.7$ eV. The experimental data is taken from Ref.\
  \onlinecite{mosnegna}.
\label{Fig:3Na7Exp}} 
\end{figure}

The results for Na$_7^-$ are shown in Fig.\ \ref{Fig:3Na7Exp}. As said
previously, in the region below $-2.5$ eV, it is difficult to compare
theory and experiment due to the great number of close lying transitions. As
in Na$_5^-$, the KS DOS describes the strongest bound large peak worst. In
this case it is already off by $0.4$ eV. In contrast, the peak position
obtained from the TDLDA excitation energies is considerably closer to the
experimental peak. It is only off by $0.1$ eV. Thus, the overestimation of the
spectrum's width by the KS DOS \cite{mosnegna, mmsk} is not observed in the
result obtained from the TDLDA calculation.  The remaining difference of $0.1$
eV between the width of the theoretical and the experimental spectrum can be
easily caused by technical aspects like the employed pseudopotential and xc
potential \cite{mmsk}. In addition, thermal effects like bond elongation and
structural isomerization can shift the obtained width by $0.1$ eV
\cite{SKTherm,mosnegna,moshhulPRL}. Considering that the experimental PES were
obtained from clusters with a temperature of around 250-300 K, the difference
between the theoretical result at zero temperature and the experimental result
is hardly surprising. At these temperatures the larger anionic sodium clusters
behave liquidlike \cite{mosnegna}. Consequently, many different ionic
configurations are present in the experiment and show up in the measured PES.

This aspect must also been kept in mind if the theoretical and experimental
results are compared in the region between $-1.3$ and $-1.7$ eV. In this
region both the zero temperature KS DOS result and the zero temperature result
from the excitation energies do not describe the measured PES very
accurately. Especially, the excitation peak at $-1.45$ eV does not fit very
well.  However, from Ref.\ \onlinecite{mosnegna} it is known that the
agreement between the experimental and the KS DOS result in this energy region
is significantly improved if different ionic structures are taken into account
via Born-Oppenheimer Langevin molecular-dynamics \cite{BarnLand}. Therefore,
one can expect that the agreement between the experimental and the TDLDA
result is also improved if different ionic structures are taken into account.
Due to the more complicated structures and the growing number of isomers the
inclusion of the temperature influence on the ionic structures of larger
clusters is much more involved than in the case of Na$_3^-$. Additionally,
combining Born-Oppenheimer Langevin molecular-dynamics with the calculation of
excitation energies is substantially more expensive than combining such a
molecular dynamics scheme with a KS DOS calculation. Thus, including thermal
effects in the present study is beyond the scope of the present work and is a
future project.

\subsection{Results for Na$_9^-$}

\begin{figure}[t]
\includegraphics[width=8.1cm]{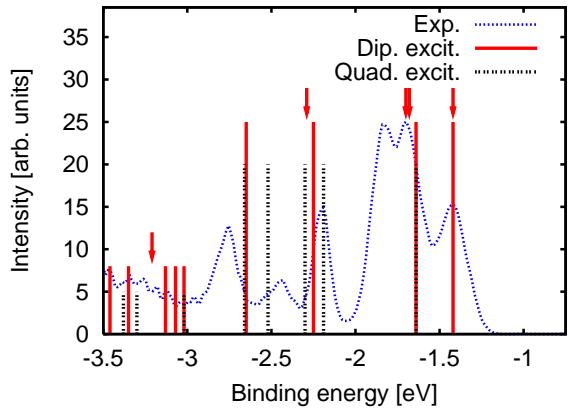}
\caption{(Color online) Same as in Fig.\ \ref{Fig:1Na3Exp} but for
  Na$_9^-$. As in Na$_7^-$, especially the stronger bound part of the spectrum
  is described more accurately by the calculated excitation energies than by
  the KS DOS. The experimental data is taken from Ref.\
  \onlinecite{mosnegna}.
\label{Fig:4Na9Exp}} 
\end{figure}

Finally, the theoretical results for Na$_9^-$ are compared in Fig.\
\ref{Fig:4Na9Exp} with the measured PES. This cluster is the first one which
has a clear peak in the range between the highest and lowest occupied KS
eigenvalue which is completely absent in the KS DOS, i.e., the experimental
PES shows six clear peaks whereas the KS DOS consists of only five peaks: the
peak around $-2.4$ eV is completely missing in the KS DOS. In addition, the
strongest bound peak in the KS DOS is off by $0.5$ eV. In other words, the KS
DOS result is inaccurate to the extent of being useless below $-2.2$ eV. As
for Na$_7^-$, the splitting of the large peak around $-1.8$ eV is reproduced
if different ionic structures are used \cite{mosnegna}.

In contrast to the KS DOS result, the PES obtained from the excitation
energies is close to the measured curve over the whole range. Below the lowest
lying peak at $-2.7$ eV the comparison is again difficult without knowing the
matrix elements mentioned above. For the two peaks at $-2.7$ and $-2.4$ eV the
theoretical values are off by $0.1$ eV.  Especially since Na$_9^-$, in
contrast to Na$_7^-$, is not a closed-shell cluster, one can expect that such
energy differences can be easily caused by ionic structure modifications
induced by finite temperatures. As expected from Ref. \onlinecite{mosnegna},
the splitting of the peak at $-1.8$ eV is also not reproduced by the zero
temperature TDLDA calculation. All in all, the experimental result in the
weaker bound part of the spectrum is described equally well by the KS DOS and
the excitation energies of the `daughter' system. However, in the stronger
bound part the time-dependent calculation yields a much more realistic
description of the PES than the KS DOS. Since this emerges as a general
observation for all systems studied in this manuscript, we discuss it on
general grounds in the following section.

\section{Summary and Conclusion}
\label{Sec:SumCon}

Using TDDFT we have calculated the excitation energies of small neutral sodium
clusters. The energies of the excited states were retrieved from the dipole
and quadrupole moments of the time-dependent density via spectral
analysis. The time-dependent density was created by an incoherent boost of all
KS ground-state orbitals and then propagated in real-space and real-time. To
discriminate between true excited state energies and energies corresponding to
energy differences between excited states we did two calculations for all
systems using different excitation energies. For comparison with measured PES
of the anionic clusters the excitation energies were calculated in the ionic
configuration of the anions.

In general, the PES for all clusters studied in this manuscript can be divided
into three parts. The first part consists of large, `weakly' bound peaks, the
second of large, `strongly' bound peaks, and finally a `less structured'
region below the lowest large experimental peak. Except for Na$_3^-$ no
comparisons between our theoretical and the experimental results can be made
in the third region.  As discussed in Subsection \ref{SubsecNa5} the main
reason for this is the missing access to the transition matrix elements
between the ground and the excited states. Since the number of excited states
can grow very rapidly, one can expect that the omission of the transition
matrix elements can cause severe problems if more complex systems are
examined. A possible way to overcome this problem is by including the
information of the matrix elements in the initial density, i.e., by creating
an initial density that only includes the states which are really excited in
the ionization process. Work in this direction is under way.

In the middle part of the spectrum the results obtained from the TDLDA
excitation energies are clearly superior to the results from the KS
DOS. Especially, the position of the strongest bound large peak is much better
reproduced by the TDLDA calculation than by the KS DOS. Thus, using the TDLDA
cures the main problem that plagues theoretical results obtained from the KS
DOS for sodium clusters, namely the prediction of a significantly too large
width of the spectrum. In addition, the PES from the TDLDA excitations can
describe an experimental peak in the PES of Na$_9^-$ which is completely
missing in the KS DOS.  The remaining differences between the experimental and
our theoretical results are all small enough to be explainable by technical
details or the finite temperature (250-300 K) of the ionic structures in the
experiment. In particular the finite temperature can be expected to be
responsible for the difference since the considered clusters behave liquidlike
at this temperature and thus, the measured PES result from many different
ionic structures / isomers which differ from the theoretical zero temperature
ground-state structures used for the calculations.

Finally, in the most weakly bound part of the spectrum we find that the TDLDA
result and the one from the KS DOS are very similar. Since the KS DOS at
finite temperature is in very good agreement with the experimental result
\cite{mosnegna}, it is extremely likely that also the TDLDA excitation energies
calculated from higher temperature ionic structures will describe the
experimental PES very well in this region.

Generally, our findings are in line with earlier results \cite{PerdNor, chong}
which report a worse agreement between the KS DOS results and the experimental
values for stronger bound levels. In addition, our results clearly show that
the agreement between the theoretical and the experimental spectrum is
considerably improved for small sodium clusters if the PES is extracted from
the true excitation energies of the `daughter' system and not the KS DOS.
This shows the importance of taking effects beyond the independent-particle
picture into account in the interpretation of photoelectron spectra.

\begin{acknowledgments}
This work is supported by the Deutscher Akademischer Austauschdienst in the
PPP Germany-Finland. We acknowledge stimulated discussions with M.\ Walter and
H.\ H\"akkinen which became possible through this support. We thank B.\ v.\
Issendorff for providing us the raw data of the measured photoelectron
spectra. S.\ K.\ also acknowledges financial support from the Deutsche
Forschungsgemeinschaft.
\end{acknowledgments}


\begin{thebibliography}{100}
\bibitem{Remark1}Ionic relaxation processes are neglected here, i.e., we assume
  that the ions do not move during the time interval between the photon
  absorption and the photoelectron detection. As a consequence, the excitation
  energies of the `daughter' system are calculated in the ionic configuration
  of the `mother' system.
\end{thebibliography}
\end{document}